\documentclass[3p,times,procedia]{elsarticle}
\usepackage[latin9]{inputenc}
\usepackage{amssymb}
\usepackage{esint}
\usepackage{bbm}
\makeatletter

\usepackage{nupha_ecrc}

\volume{00}

\firstpage{1}

\journalname{Nuclear Physics A}

\runauth{}

\jid{nupha}

\jnltitlelogo{Nuclear Physics A}

\usepackage[figuresright]{rotating}

\makeatother

\begin{document}
\begin{frontmatter}



\dochead{XXVIIIth International Conference on Ultrarelativistic Nucleus-Nucleus Collisions\\
 (Quark Matter 2019)}

\title{Chiral Anomaly, Dirac Sea and Berry monopole in Wigner Function Approach}


\author[a]{Ren-Hong Fang}
\author[b]{Jian-Hua Gao$^*$\corref{cor1}}

\cortext[cor1]{gaojh@sdu.edu.cn}

\address[a]{Institute of Particle Physics and Key Laboratory of Quark and Lepton
Physics (MOS), Central China Normal University, Wuhan 430079, China}
\address[b]{Shandong  Key Laboratory of Optical Astronomy and Solar-Terrestrial
Environment, School of Space Science and Physics, Shandong University, Weihai,
Shandong 264209, China}

\begin{abstract}
Within Wigner function formalism,  the chiral anomaly arises naturally  from the Dirac sea contribution  in un-normal-ordered Wigner function.
For massless fermions, the Dirac sea contribution behaves like a 4-dimensional or 3-dimensional Berry monopole in Euclidian momentum space, while for massive fermions, although Dirac sea still leads to the chiral anomaly but there is no Berry monopole at infrared momentum region. We discuss these points explicitly in a simple and concrete example.
\end{abstract}
\end{frontmatter}



\section{Introduction}

Under the background of the electromagnetic field, the axial vector
current of a chiral fermion system is not conserved  at the quantum
level, which is called   Adler-Bell-Jackiw anomaly or chiral anomaly.
There are many methods to study this anomaly, such as Feynmann diagram, functional integral and so on.
In recent years, there have been a considerable amount of work on the chiral kinetic theory which are devoted incorporating the chiral anomaly into the kinetic theory in a consistent way  \citep{Son:2012wh,Stephanov:2012ki,Manuel:2014dza,Chen:2012ca,Hidaka:2016yjf,Huang:2018wdl,Gao:2018wmr,Son:2012zy,Mueller:2017lzw}. Among these publications, most works connect the chiral anomaly in the chiral kinetic theory with Berry monopole in momentum space.

It has been shown in \citep{Gao:2019zhk} that  the Dirac sea or vacuum contribution from the anti-commutation relations
between antiparticle field operators in un-normal-ordered Wigner function plays a central role to generate   chiral anomaly in quantum kinetic theory both for massive and for massless fermion systems.
In this work, we will take a simple and concrete example to illustrate these points. In particular, we find
 that for massless fermion system, the chiral anomaly is associated with the singular  4-dimensional divergence
 $\partial^{\mu}[p_{\mu}\delta^{\prime}(p^{2})]$, which behaves like a 4-dimensional
 Berry monopole in Euclidian momentum space after Wick rotation, or like a 3-dimensional Berry monopole after integrating over the zero component of
 momentum. For massive fermion system, the chiral anomaly is associated with the   4-dimensional divergence
 $\partial^{\mu}[p_{\mu}\delta^{\prime}(p^{2}-m^2)]$. However it does not give rise to the singularity like 4-dimensional or 3-dimensional  Berry monopole at infrared momentum region.

\section{The divergence of axial vector current from Wigner function approach}

\label{sec:divergence}
We will take the collisionless fermion system near equilibrium under static and homogenous electromagnetic fields as  a simple and concrete example. We consider the massless fermion system first and then generalize the main results to massive case. Our starting
point is the following covariant and gauge-invariant Wigner function
for spin-1/2 fermion \citep{Vasak:1987um},
\begin{equation}
\mathcal{W}_{\alpha\beta}(x,p)= \frac{1}{(2\pi)^{4}}\int d^{4}y e^{-ip\cdot y}
\left\langle\bar{\Psi}_{\beta}\left(x+{y}/{2}\right)U\left(x+{y}/{2},x-{y}/{2}\right)
\Psi_{\alpha}\left(x-{y}/{2}\right)\right\rangle ,\label{eq:oo1}
\end{equation}
where $\langle\cdots\rangle$ represents the ensemble average, $\Psi(x)$
is the Dirac filed operator, $\alpha$, $\beta$ are Dirac spinor
indices, and $U(x+y/2,x-y/2)$ is a gauge link along a straight line
from $x-y/2$ to $x+y/2$. The dynamical equation
for $\mathcal{W}(x,p)$ is given by \citep{Vasak:1987um,Elze:1986qd,Elze:1989un},
\begin{equation}
\label{Wig-eqn}
\gamma^{\mu}\bigg(p_{\mu}+\frac{i}{2}\nabla_{\mu}\bigg)\mathcal{W}(x,p)=0,\label{eq:oo2}
\end{equation}
where $\nabla_{\mu}=\partial_{\mu}^{x}-QF_{\mu\nu}\partial_{p}^{\nu}$.
It should be noted that there is no normal ordering in the Wigner matrix above.
This plays a central role to give rise to the chiral anomaly in the following.
Since $\mathcal{W}(x,p)$ is a $4\times4$ matrix, we can decompose
it by the 16 independent $\Gamma$-matrices,
\begin{equation}
\mathcal{W}=\frac{1}{4}\bigg(\mathcal{F}+i\gamma^{5}\mathcal{P}+\gamma^{\mu}\mathcal{V}_{\mu}+\gamma^{5}\gamma^{\mu}\mathcal{A}_{\mu}+\frac{1}{2}\sigma^{\mu\nu}\mathcal{S}_{\mu\nu}\bigg).\label{eq:oo3}
\end{equation}
The vector current $J_{V}^{\mu}(x)$ and axial vector current $J_{A}^{\mu}(x)$
can be expressed as the 4-momentum integration of $\mathcal{V}^{\mu}$
and $\mathcal{A}^{\mu}$. We can expand $\mathcal{V}^{\mu}$ and $\mathcal{A}^{\mu}$
order by order in $\hbar$ as
\begin{eqnarray}
\mathcal{V}^{\mu} & = & \mathcal{V}_{(0)}^{\mu}+\hbar\mathcal{V}_{(1)}^{\mu}+\hbar^{2}\mathcal{V}_{(2)}^{\mu}+\cdots,\label{eq:ap35}\\
\mathcal{A}^{\mu} & = & \mathcal{A}_{(0)}^{\mu}+\hbar\mathcal{A}_{(1)}^{\mu}+\hbar^{2}\mathcal{A}_{(2)}^{\mu}+\cdots.\label{eq:ap36}
\end{eqnarray}
As mentioned above, we will consider the specific fermion system near equilibrium. Hence we choose the zeroth order solutions $\mathcal{V}_{(0)}^{\mu}$ and $\mathcal{A}_{(0)}^{\mu}$
as equilibrium distribution in free field theory \citep{Fang:2016vpj,Gao:2012ix,Gao:2019zhk}
\begin{eqnarray}
\mathcal{V}_{(0)}^{\mu} & = & p^{\mu}\delta(p^{2})\sum_{s}(\mathcal{Z}_{s}^{\mathrm{n}}+\mathcal{Z}^{\mathrm{v}}),\label{eq:h2}\\
\mathcal{A}_{(0)}^{\mu} & = & p^{\mu}\delta(p^{2})\sum_{s}s\mathcal{Z}_{s}^{\mathrm{n}},\label{eq:s2-1}
\end{eqnarray}
where $\mathcal{Z}_{s}^{\mathrm{n}},\mathcal{Z}^{\mathrm{v}}$ are given by
\begin{eqnarray}
\mathcal{Z}_{s}^{\mathrm{n}} & = & \frac{2}{(2\pi)^{3}}[\theta(u\cdot p)n_{F}(u\cdot p-\mu_{s})+\theta(-u\cdot p)n_{F}(-u\cdot p+\mu_{s})],\label{eq:h3}\\
\mathcal{Z}^{\mathrm{v}} & = & -\frac{2}{(2\pi)^{3}}\theta(-u\cdot p).\label{eq:h1}
\end{eqnarray}
Here $n_{F}(x)=1/(e^{\beta x}+1)$ is the Fermi-Dirac distribution,
$\beta=1/T$ is the inverse temperature, $u^{\mu}$ is the fluid velocity,
and $\mu_{s}=\mu+s\mu_{5}$ with $s=\pm1$ is the chemical potential
for righthand/lefthand fermions respectively. For simplicity, we will
assume that there is no vorticity in the system, i.e. the fluid velocity
$u^{\mu}$ is uniform. Note that $\mathcal{Z}^{\mathrm{v}}$ is the
vacuum term or Dirac sea term which comes from the anticommutation of creation and
annihilation operators of antifermions as described in \citep{Gao:2019zhk,Sheng:2017lfu,Sheng:2018jwf}.
This term is universal and does not depend on the specific distribution $n_{F}(x)$ at all.
From the dynamical equation of Wigner function at order $\hbar$,
one can obtain $\mathcal{A}_{(1)}^{\mu}$ near equilibrium
\begin{equation}
\mathcal{A}_{(1)}^{\mu}=Q\tilde{F}^{\mu\nu}p_{\nu}\delta^{\prime}(p^{2})\sum_{s}(\mathcal{Z}_{s}^{\mathrm{n}}+\mathcal{Z}^{\mathrm{v}}),\label{eq:qq2}
\end{equation}
where $\tilde{F}^{\mu\nu}=\frac{1}{2}\epsilon^{\mu\nu\rho\sigma}F_{\rho\sigma}$.
The equation satisfied by $\mathcal{A}^{\mu}$ can be obtained  from Eq.(\ref{Wig-eqn}),
\begin{equation}
(\partial_{\mu}^{x}-QF_{\mu\nu}\partial_{p}^{\nu})\mathcal{A}^{\mu}=0.\label{eq:qq4}
\end{equation}
The 4-momentum integration of Eq. (\ref{eq:qq4}) gives
\begin{equation}
\partial_{\mu}J_{A}^{\mu}=QF_{\mu\nu}\int d^{4}p\partial_{p}^{\nu}\mathcal{A}_{(1)}^{\mu}\equiv-\frac{Q^{2}}{8\pi^{2}}F_{\mu\nu}\tilde{F}^{\mu\nu}\times(C_{\mathrm{n}}+C_{\mathrm{v}}),\label{eq:h3-1}
\end{equation}
where $C_{\mathrm{n}},C_{\mathrm{v}}$ are defined as
\begin{eqnarray}
C_{\mathrm{n}} & = & -2\pi^{2}\int d^{4}p\partial_{p}^{\rho}\bigg[p_{\rho}\delta^{\prime}(p^{2})\sum_{s}\mathcal{Z}_{s}^{\mathrm{n}}\bigg],\label{eq:h6}\\
C_{\mathrm{v}} & = & -4\pi^{2}\int d^{4}p\partial_{p}^{\rho}[p_{\rho}\delta^{\prime}(p^{2})\mathcal{Z}^{\mathrm{v}}].\label{eq:h7}
\end{eqnarray}
 All discussions above is for massless fermion system.
For massive fermion case, it turns out  that we can just set $\mu_{s}=\mu$, $\mu_{5}=0$
and change the on-shell condition in Eq. (\ref{eq:h6}, \ref{eq:h7})
as $\delta^{\prime}(p^{2})\rightarrow\delta^{\prime}(p^{2}-m^{2})$,
and Eq. (\ref{eq:h3-1}) becomes
\begin{equation}
\partial_{\mu}J_{A}^{\mu}=-2m\int d^{4}p\,\mathcal{P}-\frac{Q^{2}}{8\pi^{2}}F_{\mu\nu}\tilde{F}^{\mu\nu}\times(C_{\mathrm{n}}+C_{\mathrm{v}}),\label{eq:h8}
\end{equation}
where $\mathcal{P}$ is the pseudoscalar component of Wigner function
in Eq. (\ref{eq:oo3}).

\section{Chiral anomaly  for massless fermion system }

For massless fermion case, since $\mathcal{Z}_{s}^{\mathrm{n}}$ vanish rapidly  at infinity in the phase space,
$C_{\mathrm{n}}$ must be zero.  However the Dirac sea term $C_{\mathrm{v}}$ can keeps nonzero at $p_0=-\infty$ and could contribute non-zero value
\begin{eqnarray}
C_{\mathrm{n}} & = & 0,\label{eq:h11}\\
C_{\mathrm{v}} & = & \frac{1}{\pi}\int d^{4}p\partial^{\mu}[p_{\mu}\theta(-p^{0})\delta^{\prime}(p^{2})]
=\frac{1}{2\pi}\int d^{4}p\partial^{\mu}[p_{\mu}\delta^{\prime}(p^{2})].\label{eq:h12}
\end{eqnarray}
We can calculate this integral by two methods. First we can use the regularization
\begin{eqnarray}
\delta'(x)=\frac{1}{\pi}\textrm{Im}\frac{1}{(x+i\epsilon)^2},
\end{eqnarray}
followed by  Wick rotation and  obtain
\begin{equation}
C_{\mathrm{v}}  =  \frac{1}{2\pi^2}\textrm{Im}\int d^{4}p\, \partial^{\mu}\left[\frac{p_{\mu}}{(p^2+i\epsilon)^2}\right]
=\frac{1}{2\pi^2}\int d^{4}p_E\, \partial_{\mu}\left(\frac{p^{\mu}_E}{p_E^4}\right)
=1,
\end{equation}
where we have used the Gauss theorem in 4-dimensional momentum space or the identity $\partial_{\mu}({p^{\mu}_E}/{p_E^4})=2\pi^2 \delta^4(p_E)$.
It is obvious that $p_{\mu}\delta^{\prime}(p^{2})$
plays the role of the Berry curvature of a 4-dimensional monopole in Euclidean
momentum space, which was  pointed out in Ref. \citep{Chen:2012ca}.

We can also calculate the integral by brute force. After integrating over the zero component of momentum and keeping the non-vanishing term,  we obtain
\begin{equation}
C_{\mathrm{v}} = \int \frac{d^3{\bf p}}{2\pi}{\mathbf \partial}_{\bf p}\cdot \left( \,
\frac{\hat{\bf p}}{2{\bf p}^2}\right)
=1,
\end{equation}
where we have used the  Gauss theorem in 3-dimensional momentum space or the identity
${\partial}_{\bf p}\cdot (\hat{\bf p}/2{\bf p}^2)=2\pi \delta^3({\bf p})$. Here $\hat{\bf p}/2{\bf p}^2$
is just the usual Berry curvature in 3-dimensional momentum space.  For massless fermion system, we note that only vacuum or  Dirac sea term contributes
to the chiral anomaly in form of 4-dimensional or 3-dimensional Berry monopole.
\section{Chiral anomaly  for massive fermion system }

For massive fermion system, it turns out that the coefficients $C_{\mathrm{n}},C_{\mathrm{v}}$ can be obtained  by replacing the on-shell condition
$\delta^{\prime}(p^{2})$ with $\delta^{\prime}(p^{2}-m^{2})$. Similar to the massless fermion system, $C_{\mathrm{n}}$ always vanishes for
normal distribution, while $C_{\mathrm{v}}$ is given by
\begin{eqnarray}
C_{\mathrm{v}} & = & \frac{1}{\pi}\int d^{4}p\partial^{\mu}[p_{\mu}\theta(-p^{0})\delta^{\prime}(p^{2}-m^2)]
=\frac{1}{2\pi}\int d^{4}p\partial^{\mu}[p_{\mu}\delta^{\prime}(p^{2}-m^2)].\label{eq:h12}
\end{eqnarray}
Again we can calculate this integral by Wick rotation
\begin{equation}
C_{\mathrm{v}}
=\frac{1}{2\pi^2}\textrm{Im}\int d^{4}p\, \partial^{\mu}\left[\frac{p_{\mu}}{(p^2-m^2+i\epsilon)^2}\right]
=\frac{1}{2\pi^2}\int d^{4}p_E\, \partial_{\mu}\left[\frac{p^{\mu}_E}{(p_E^2+m^2)^2}\right]
=1,
\end{equation}
or directly integrate over $p_0$
\begin{equation}
C_{\mathrm{v}} = \int \frac{d^3{\bf p}}{2\pi}{\mathbf \partial}_{\bf p}\cdot
 \left[ \, \frac{\hat{\bf p}}{2({\bf p}^2+m^2)}\right]
=1,
\end{equation}
where we have used Gauss theorem in 4-dimensional and 3-dimensional momentum space, respectively.
We can notice that there is no singular Berry curvature of a 4-dimensional or 3-dimensional monopole in Euclidean
momentum space here due to the presence of finite mass.

The chiral anomaly derived from Dirac sea contribution for massless or massive case is universal and
 independent of the phase space  normal distribution function at zero momentum.

\textit{Acknowledgments}.
This work was supported in part by the National Natural Science Foundation of China under
Nos. 11847220 and 11890713. R.-H. F. thanks for the hospitality of Institute of Frontier and Interdisciplinary Science at Shandong University (China) where he is currently visiting.

 \bibliographystyle{elsarticle-num}
\bibliography{ref-1}



\end{document}